\documentclass[lettersize,journal]{IEEEtran}

\usepackage{myStyle}
\usepackage{subcaption}
\usepackage{setspace}
\usepackage{pgfornament}
\usepackage{eso-pic}
\AddToShipoutPictureBG*{%
	\AtPageUpperLeft{%
		\setlength\unitlength{1in}%
		\hspace*{\dimexpr0.5\paperwidth\relax}
		\makebox(0,-0.5)[c]{\begin{tabular}{c c}
				Max van Haren \textit{et al}. Local Rational Modeling for Identification Beyond the Nyquist Frequency: \\ Applied to a Prototype Wafer Stage, 
				In preparation for journal submission, uploaded \today 
		\end{tabular}}
}}
\newcommand{\citet}{\cite}
\newcommand{\citep}{\cite}
\begin{document}
	\title{Local Rational Modeling for Identification Beyond the Nyquist Frequency: Applied to a Prototype Wafer Stage}

\author{Max van Haren, Lennart Blanken, Koen Classens, and Tom Oomen
	\thanks{Submitted for review on Oktober 3, 2024. This work is part of the research programme VIDI with project number 15698, which is (partly) financed by the Netherlands Organisation for Scientific Research (NWO). In addition, this research has received funding from the ECSEL Joint Undertaking under grant agreement 101007311 (IMOCO4.E). The Joint Undertaking receives support from the European Union Horizon 2020 research and innovation programme. (\textit{Corresponding author: Max van Haren.})}
	\thanks{Max van Haren is with the Control Systems Technology Section, Department of Mechanical Engineering, Eindhoven University of Technology, 5600 MB Eindhoven, The Netherlands (e-mail: m.j.v.haren@tue.nl).}%
	\thanks{Lennart Blanken is with Sioux Technologies, 5633 AA Eindhoven, The Netherlands. He is also with the Control Systems Technology Section, Department of Mechanical Engineering, Eindhoven University of Technology, 5600 MB Eindhoven, The Netherlands (e-mail: l.l.g.blanken@tue.nl).}%
	\thanks{Koen Classens is with the Control Systems Technology Section, Department of Mechanical Engineering, Eindhoven University of Technology, 5600 MB Eindhoven, The Netherlands (e-mail: k.h.j.classens@tue.nl).}%
	\thanks{Tom Oomen is with the Control Systems Technology Section, Department of Mechanical Engineering, Eindhoven University of Technology, 5600 MB Eindhoven, The Netherlands. He is also with the Delft Center for Systems and Control, Delft University of Technology, 2628 CD Delft, The Netherlands (e-mail: t.a.e.oomen@tue.nl).}%
}	
	\maketitle
	 \pagestyle{empty}
	 \thispagestyle{empty}
	\begin{abstract}
		Fast-rate models are essential for control design, specifically to address intersample behavior. The aim of this \manuscript is to develop a frequency-domain non-parametric identification technique to estimate fast-rate models of systems that have relevant dynamics and allow for actuation above the Nyquist frequency of a slow-rate output. Examples of such systems include vision-in-the-loop systems. Through local rational models over multiple frequency bands, aliased components are effectively disentangled, particularly for lightly-damped systems. The developed technique accurately determines non-parametric fast-rate models of systems with slow-rate outputs, all within a single identification experiment. Finally, the effectiveness of the technique is demonstrated through experiments conducted on a prototype wafer stage used for semiconductor manufacturing.
	\end{abstract}
	
\section{Introduction}
\label{sec:introduction}
\IEEEPARstart{S}{ystems} with actuation and dynamics that exceed the Nyquist frequency of the sensor are becoming increasingly more common in mechatronics, for example in hard disk drives \citep{Atsumi2010} and vision-in-the-loop applications \citep{Hutchinson1996}. The Nyquist-Shannon sampling theorem \citep{Shannon1949} implies that these systems are usually identified up to the Nyquist frequency of the slow-rate sensor. On the other hand, fast-rate models are often necessary for tasks such as control design \citep{Atsumi2010,Mae2023} and intersample performance assessment \citep{Oomen2007}.

Control design and performance evaluation of linear time-invariant (LTI) systems often involves non-parametric frequency-domain models. Techniques such as manual loop-shaping \citep{Schmidt2020} and parametric system identification \citep{Pintelon1994} commonly employ non-parametric frequency-domain models. Frequency Response Functions (FRFs) are widely used to represent systems in the frequency domain, and can be directly identified from input-output data, providing a quick, accurate, and cost-effective solution \citep{Pintelon2012,Oomen2018a}. Furthermore, FRFs enable the direct analysis of stability, performance, and robustness \citep{Skogestad2005}.

Recently, more advanced FRF identification techniques have been developed, including the Local Polynomial Modeling (LPM) method \citep{Schoukens2009,Pintelon2010a}. LPM essentially estimates a polynomial model in a local frequency window using a least-squares cost function. LPM generally leads to an improved estimate of the FRF, which is mainly enabled by the concurrent estimation and suppression of transient contributions. Following the LPM method, the Local Rational Modeling (LRM) technique has been developed. Unlike LPM, LRM estimates a rational model within a local frequency window \citep{McKelvey2012, Voorhoeve2018}, which is shown to be more effective for lightly-damped resonant dynamics \citep{Geerardyn2016}.


Irrespective of the identification approach, identifying fast-rate models using slow-rate outputs is challenging due to aliasing. Aliasing occurs when a signal is sampled at a rate that is insufficient to capture the fast-rate dynamics of a system, preventing the unique recovery of the associated fast-rate model \citep{Astrom2011}.

Substantial research has been done on identification of fast-rate models using slow-rate outputs, with a primary focus on continuous-time and multirate parametric system identification. First, continuous-time system identification identifies a continuous-time parametric model using input-output data \citep{Unbehauen1990}. These methods typically constrain the input signal, such as zero-order hold or band-limited signals \citep{Ljung2009, Ehrlich1989}. Second, parametric identification of fast-rate models using multirate data has been developed, including impulse response \citep{Ding2004}, ARX \citep{Ding2008a,Ding2011} and output-error \citep{Zhu2009} model estimation. Furthermore, state-space models of multirate systems are generally identified using the lifting technique \citep{Li2001,Ding2005}. All these methods focus on parametric models, and in addition require intersample assumptions on the input signal and do not take full advantage of fast-rate inputs, thereby failing to disentangle aliased components.

A recent study introduced a novel non-parametric approach for identifying fast-rate models beyond the Nyquist of a slow-rate output, where aliased contributions are disentangled by assuming locally smooth behavior of the FRF \citep{VanHaren2023a}. The resulting method identifies FRFs in a single identification experiment beyond the Nyquist frequency. However, the local smoothness assumption is not capable of modeling resonant dynamics accurately \citep{McKelvey2012}.

Although methods for identification beyond the Nyquist frequency of slow-rate outputs have been developed, there is a need for an efficient and systematic methodology for single-experiment FRF identification of fast-rate models, that disentangles aliased components with broadband input signals. In this \manuscript, fast-rate models are identified with broadband excitation signals and slow-rate outputs, where through the use of local rational models over multiple frequency bands, aliased components are disentangled from each other. The use of local models is at the foundation of modern FRF identification for LTI single-rate systems, such as the LPM and LRM techniques. In fact, both LPM and LRM for LTI single-rate systems are recovered as a special case of the developed framework. The key contributions of this \manuscript include the following.
\begin{itemize}
\item[C1] Formulation of a non-convex optimization for FRF identification beyond the Nyquist frequency, based on local rational modeling across multiple frequency bands to disentangle aliased components.
\item[C2] A solution approach through an appropriately weighted linear least-squares criterion, which has a closed-form minimizer. Furthermore, the accuracy of the weighted cost is improved through the use of iterative reweighted solutions.
\item[C3] Validation of the developed framework on an experimental prototype wafer stage used in for semiconductor manufacturing.
\end{itemize}
The approach in \citet{VanHaren2023a} is recovered as a special case of the developed framework. In contrast to \citet{VanHaren2023a}, the method presented in this \manuscript is particularly suitable for lightly-damped resonant dynamics, which is validated using experimental results.

\paragraph*{Notation}
Fast-rate signals are denoted by subscript $h$, and slow-rate signals with subscript $l$, which have been downsampled by a factor $\fac\in\mathbb{Z}_{>0}$, with integers $\mathbb{Z}$. Fast-rate and slow-rate signals consists of respectively $N$ and $M=\frac{N}{\fac}$ data points. The $N$-points and $M$-points Discrete Fourier Transform (DFT) for finite-time fast-rate and slow-rate signals is given by
\begin{equation}
	\label{eq:DFT3}
		\begin{aligned}
			X_h(k) &= \sum_{\dt=0}^{N-1} x_h(\dt) e^{-j\omega_k \dt \tsh}, \\
			X_l(k) &= \sum_{\ldt=0}^{M-1} x_l(\ldt) e^{-j \omega_k \ldt \tsl } \\
			&= \sum_{\dt=0}^{M-1} x_h(\dt\fac) e^{-j \omega_k \dt \tsl},
		\end{aligned}
\end{equation}
with respectively sampling times $\tsh$ and $\tsl=\fac\tsh$, discrete-time indices for fast-rate signals $\dt\in \mathbb{Z}_{[0,N-1]}$ and slow-rate signals $\ldt\in \mathbb{Z}_{[0,M-1]}$, and frequency bin $k\in\mathbb{Z}_{[0,N-1]}$, that relates to the frequency grid
\begin{equation}
	\label{eq:omegak}
	\omega_k=\frac{2\pi k}{N \tsh}=\frac{2\pi k}{M \tsl}.
\end{equation}
The complex conjugate of $A$ is denoted as $\overline{A}$ and the complex conjugate transpose as $A^H$. The expected value of a random variable $X$ is given by $\mathbb{E}\left\{X\right\}$.
\section{Problem Formulation}
In this section, a motivating application and the identification setting are shown for identification beyond the Nyquist frequency of slow-rate outputs. Finally, the problem treated in this \manuscript is defined.
\label{sec:pdef}
\subsection{Motivating Application}
\begin{figure}[tb]
	\centering
	\includegraphics[width=0.99\linewidth]{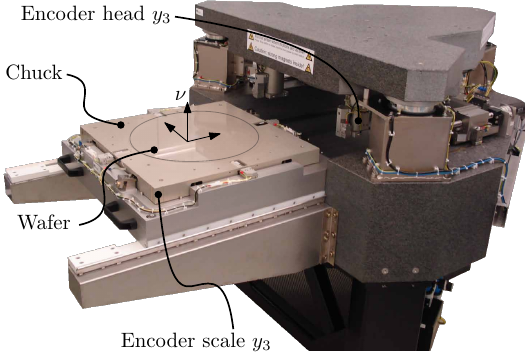}
	\caption{Photograph of experimental setup, containing the wafer stage.}
	\label{fig:ExperimentalSetup}
\end{figure}

\begin{figure}[tb]
	\centering
	\includegraphics[width=0.99\linewidth]{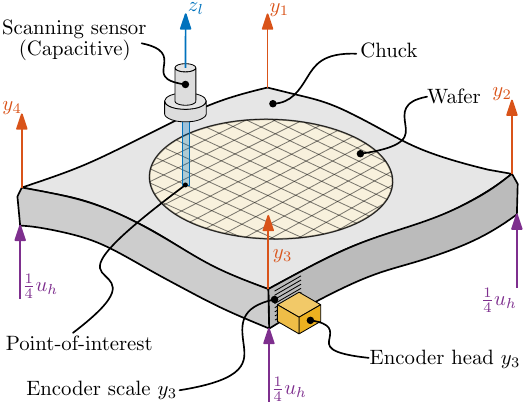}
	\caption{Schematic overview of experimental setup, where the fast-rate input $u_h$ is distributed over the four corners of the chuck. The outputs used for feedback $y_i \; \forall i \in \{1,2,3,4\}$ are measured using encoder scales and heads, which is schematically depicted for $y_3$. The performance output $z_l$ is measured using an additional capacitive scanning sensor, which is suspended less than 1 mm above the wafer and sampled at a reduced sampling rate.}
	\label{fig:ExperimentalSetup2}
\end{figure}

The problem addressed in this \manuscript is directly motivated by the considered prototype wafer stage in \figRef{fig:ExperimentalSetup}, which is used in semiconductor manufacturing. Specifically, the Over Actuated Test-rig (OAT) is a prime example of a mechatronic system with a slow-rate output. 
The objective of the OAT is to accurately control the vertical position $\nu$ of the point-of-interest on the wafer, which is the point on the wafer during lithographic exposure. Directly measuring the vertical displacement of the point-of-interest on the wafer is not possible using linear encoders. The chuck of the OAT has internal lightly-damped structural modes, and hence, measuring the vertical displacement on the outside of the chuck does not coincide with the vertical displacement at the point-of-interest on the wafer \citep{Oomen2015,Heertjes2020}. Therefore, an external capacitive sensor that directly measures the point-of-interest is employed, as denoted by scanning sensor in \figRef{fig:ExperimentalSetup2}. The external capacitive sensor is sampled at a reduced sampling rate compared to the actuators.
The OAT, characterized by its slow-rate sensor and lightly-damped resonant behavior, directly motivates the need for rational identification techniques beyond the Nyquist frequency.



\subsection{Identification Setting}
\begin{figure}[tb]
	\centering
	\includegraphics{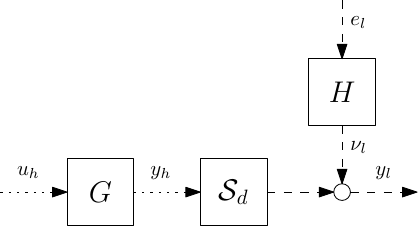}
	\caption{
Open-loop identification setting considered. The fast-rate system $G$ with fast-rate input $u_h$ and output $y_h$ have sampling times \tsh. The slow-rate output $y_l$ is downsampled and disturbed with measurement noise, i.e., $y_l=\mathcal{S}_dy_h+He_l$, and has sampling time $\tsl=\fac\tsh$.
}
	\label{fig:IdentificationSetting}
\end{figure}

The goal is to identify a non-parametric FRF of fast-rate system $G$ using slow-rate outputs $y_l$ and fast-rate inputs $u_h$, as shown in \figRef{fig:IdentificationSetting}.
The set of systems $G$ that is considered is described by the LTI discrete-time rational transfer function
\begin{equation}
	\label{eq:RationalSystem}
	\begin{aligned}
		G\left(q\right) = \frac{B\left(q\right)}{A\left(q\right)} = \frac{\sum_{i=0}^{n_b}b_i q^{-i}}{\sum_{i=0}^{n_a}a_i q^{-i}},
	\end{aligned}
\end{equation}
where $q^{-1}$ denotes the lag operator $q^{-1}x(\dt) = x(\dt-1)$. The fast-rate output $y_h$ of system $G$ under input $u_h$ is given by
\begin{equation}
	\label{eq:fastSampledOutput}
	\begin{aligned}
		y_h(\dt) = G\left(q\right) u_h\left(\dt\right) + t(\dt),
	\end{aligned}
\end{equation}
with transient contribution $t\left(\dt\right)$, which includes leakage and occurs due to finite-length signals \citep{Schoukens2009, Voorhoeve2018}. Taking the DFT on both sides of \eqref{eq:fastSampledOutput} results in
\begin{equation}
	\label{eq:FastSampledOutput}
	\begin{aligned}
		Y_h(k) = G\left( \freq_k\right) U_h(k) + T\left( \freq_k\right),
	\end{aligned}
\end{equation}
with generalized frequency variable $\freq_k = e^{-j\omega_k \tsh}$ for discrete-time systems, and transient contribution $T\left( \freq_k\right)$. Since $G$ is assumed to be LTI and described according to \eqref{eq:RationalSystem}, the fast-rate output $Y_h(k)$ is only influenced by a single frequency of $U_h(k)$, commonly denoted by the frequency-separation principle. The fast-rate output is downsampled, as shown in \figRef{fig:IdentificationSetting}, into
\begin{equation}
	\label{eq:GeneralFDomainDownsampledOutput}
	\begin{aligned}
		Y_l(k) = \mathcal{S}_d Y_h(k)+V_l(k),
	\end{aligned}
\end{equation} 
with noise $V_l(k)=H(\freq_k)E(k)$, where $E(k)$ is zero-mean independent and identically distributed noise. The transient of the noise system $H$ is typically neglected, since it is negligible compared to its steady-state response $He_l$ \citep[Section~6.7.3.4]{Pintelon2012}. The downsampling operation in \eqref{eq:GeneralFDomainDownsampledOutput} is equal to the time-domain operation $y_l(\ldt)=y_h(\dt\fac)+v_l(m)$. The DFT of the slow-rate output is found by expanding the downsampling operation in \eqref{eq:GeneralFDomainDownsampledOutput} \citep{Vaidyanathan1993}, i.e.,
\begin{equation}
	\label{eq:FDomainSlowSampledOutput1}
	\begin{aligned}
		Y_l(k)=\frac{1}{\fac} \sum_{f=0}^{\fac-1} Y_h(k+fM)+V_l(k).
	\end{aligned}
\end{equation}
By substituting the fast-rate output from \eqref{eq:FastSampledOutput} into \eqref{eq:FDomainSlowSampledOutput1}, the slow-rate output results in
\begin{equation}
	\label{eq:DFT_YL}
	\begin{aligned}
		Y_l(k)\!=\!\!\frac{1}{F} \!\!\sum_{f=0}^{\fac-1}\!\!\Big(\!G\!\left(\freq_{k+fM}\right) U_h\!(k\!+\!\!fM)\!+\!\!T_G\left(\freq_{k+fM}\right)\!\!\Big)\!\!+\!\!V_l(k).
	\end{aligned}
\end{equation}
\subsection{Problem Definition}
The downsampler results in the DFT of the output \eqref{eq:DFT_YL} being affected by \fac contributions from the system $G(\freq_{k+fM})$ and transient $T(\freq_{k+fM})$. Therefore, the fast-rate FRF $G(\freq_{k})$ can in general not be uniquely identified beyond the Nyquist frequency of the slow-rate using fast-rate inputs $u_h$.

The problem considered is as follows. Given fast-rate inputs $u_h$ and slow-rate outputs $y_l$ from the system $\mathcal{S}_dG$ in \figRef{fig:IdentificationSetting}, identify a fast-rate FRF $\widehat{G}(\freq_{k})$ for the frequencies $\freq_{k} \; \forall k\in\mathbb{Z}_{[0,N-1]}$, i.e., up to the fast sampling frequency $\fsh=\frac{1}{\tsh}$. Throughout the \manuscript, requirements on the fast-rate input signal $u_h$ are investigated as well.



\section{Local Rational Modeling Beyond the Nyquist Frequency}
\label{sec:NLMethod}
In this section, local rational models across multiple frequency bands are developed to identify fast-rate models beyond the Nyquist frequency of a slow-rate sensor under broadband excitation, leading to contribution C1.

The rational description of $G$ in \eqref{eq:RationalSystem} motivates parameterizing a model within the local frequency window $r\in\mathbb{Z}_{[-\wsize,\wsize]}$, with $2\wsize+1$ the window size, as
\begin{equation}
	\label{eq:localPlant}
	\begin{aligned}
		\widehat{G}(\freq_{k+r+fM}) = \frac{n(\freq_{k+r+fM})}{d(\freq_{k+r+fM})},
	\end{aligned}
\end{equation}
for all $f\in\mathbb{Z}_{[0,\fac-1]}$, where
\begin{equation}
	\label{eq:ndParameterization}
	\begin{aligned}
		n(\freq_{k+r+fM}) &= \widehat{G}(\freq_{k+fM}) +\sum_{s=1}^{R_n}n_s(k+fM)r^s, \\
		d(\freq_{k+r+fM}) & = 1+\sum_{s=1}^{R_d}d_s(k+fM)r^s.
	\end{aligned}
\end{equation}
Similarly, the transient $T$ is parameterized as
\begin{equation}
	\label{eq:localTransient}
	\begin{aligned}
		\widehat{T}(\freq_{k+r+fM}) = \frac{m(\freq_{k+r})}{d(\freq_{k+r+fM})},
	\end{aligned}
\end{equation}
where the same denominator is used as for the system $G$, since $G$ and $T$ share the same poles \citep{McKelvey2012}. Solely the contribution of the transient in the slow-rate output $Y_l$ is of interest, and hence, the numerator of the transient $m(\freq_{k+r})$ is modeled at the slow-rate and independently of the frequency band $f$, i.e.,
\begin{equation}
	\begin{aligned}
		m(\freq_{k+r}) = \widehat{T}(\freq_{k}) +\sum_{s=1}^{R_m}m_s(k)r^s,
	\end{aligned}
\end{equation}
which allows to suppress the transient contribution.

The slow-rate output in window $k+r$ is estimated using the local models of the system and transient $\widehat{G}$ and $\widehat{T}$, given the input signal $U(k+r)$, as
\begin{equation}
	\label{eq:Ylapprox}
	\begin{aligned}
		\widehat{Y}_l(k+r) = \frac{1}{\fac}\sum_{f=0}^{\fac-1}&\Big(\widehat{G}\left(\freq_{k+r+fM}\right)U(k+r+fM)\\
		+&\widehat{T}\left(\freq_{k+r+fM}\right)\Big).
	\end{aligned}
\end{equation}
The local parameterization of the system and transient in \eqref{eq:localPlant} and \eqref{eq:localTransient} leads to the estimated output
\begin{equation}
	\label{eq:NonLinearAdditiveYlApprox}
	\begin{aligned}
		&\widehat{Y}_l(k+r) = \frac{1}{\fac}\sum_{f=0}^{\fac-1}\frac{\widehat{T}(\freq_k)+\sum_{s=1}^{R_m}m_s(k)r^s}{1+\sum_{s=1}^{R_d}d_s(k+fM)r^s)} \\
		+&\frac{1}{\fac}\sum_{f=0}^{\fac-1}\frac{\widehat{G}(\freq_{k+fM})+\sum_{s=1}^{R_n}n_s(k+fM)r^s}{1+\sum_{s=1}^{R_d}d_s(k+fM)r^s}U(k\!+\!r\!+\!fM).
	\end{aligned}
\end{equation}
The parameters are gathered in a vector $\widehat{\Theta}(k)\in\mathbb{C}^{1\times F(R_n+1+R_d)+R_m+1}$, i.e.,
\begin{equation}
	\label{eq:NonLinearParameterVector}
	\begin{aligned}
		\widehat{\Theta}(k) = \begin{bmatrix}
			\theta_{\widehat{G}} \!\! & \theta_{N} & \widehat{T}(\freq_{k})\!\! & m_1(k) & \!\!\!\!\!\cdots\!\!\!\!\! & m_{R_m}(k)\!\! &  \theta_{D}
		\end{bmatrix},
	\end{aligned}
\end{equation}
where
\begin{align}
	\theta_{\widehat{G}} &= \frac{1}{\fac} \begin{bmatrix}
		\widehat{G}(\freq_{k}) & \widehat{G}(\freq_{k+M}) & \cdots & \widehat{G}(\freq_{k+(\fac-1)M})
	\end{bmatrix}, \nonumber \\
	\theta_N &=  \frac{1}{\fac} \begin{bmatrix}
		\theta_{n_1} & \theta_{n_2} & \cdots & \theta_{n_{R_n}}
	\end{bmatrix},  \label{eq:NonLinearParameters} \\ 
	\theta_D &= \begin{bmatrix}
		\theta_{d_1} & \theta_{d_2} & \cdots & \theta_{d_{R_d}}
	\end{bmatrix}, \nonumber
\end{align}
and
\begin{equation}
	\begin{aligned}
			\theta_{n_i} &=\begin{bmatrix}
			n_i({k}) & n_i({k+M}) & \cdots & n_i({k+(\fac-1)M})
		\end{bmatrix}, \\
		\theta_{d_i} &= \begin{bmatrix}
		d_i({k}) & d_i({k+M}) & \cdots & d_{i}({k+(\fac-1)M})
	\end{bmatrix}.
	\end{aligned}
\end{equation}
The decision parameters $\widehat{\Theta}$ are determined by optimizing an objective function, i.e.,
\begin{equation}
	\label{eq:Minimization}
	\begin{aligned}
		\min_{\widehat{\Theta}(k)} J(\widehat{\Theta}(k)).
	\end{aligned}
\end{equation}
As objective function, the least-squares residual between estimated and measured output $Y_l$ within the local frequency window $k+r$ is given by
\begin{equation}
	\label{eq:CostFunction}
	\begin{aligned}
		J_{LS}\left(\widehat{\Theta}\left(k\right)\right)= \sum_{r=-\wsize}^{\wsize}\left\|
		Y_l(k+r) - \widehat{Y}_l\left(k+r,\widehat{\Theta}\left(k\right)\right)
		\right\|_2^2.
	\end{aligned}
\end{equation}
\begin{remark}
	LRM for single-rate LTI systems with a non-linear cost function \citep{Voorhoeve2018} is recovered as a special case of the framework by setting $\fac=1$. Furthermore, if $\fac=1$ and $d(\freq_{k+r})=1$, i.e., $R_d = 0$, LPM for single-rate LTI systems \citep{Schoukens2009,Pintelon2010a} is recovered.
\end{remark}

Optimizing cost function \eqref{eq:CostFunction} is challenging because \begin{enumerate*} \item it involves a summation due to the downsampling operation, and \item the system and transient are rationally parameterized as shown in \eqref{eq:NonLinearAdditiveYlApprox}. \end{enumerate*} Given the rational model structure and the downsampling operation, the cost function in \eqref{eq:CostFunction} is non-linear with respect to the parameters $\widehat{\Theta}(k)$. As a result, it is generally non-convex and does not have a closed-form solution.

\section{FRF Identification Beyond the Nyquist Frequency with Local Rational Models}
\label{sec:LinearMethod}
In this section, a solution approach for unique and convex identification of local rational models beyond the Nyquist frequency is presented, leading to contribution C2. The key idea is to appropriately weight the non-linear cost \eqref{eq:CostFunction}, leading to a linear least-squares criterion. The unique existence of the closed-form solution is guaranteed through design conditions on the input and local models.
Additionally, the closed-form solution enables to approximate the variance of the FRF. Subsequently, the accuracy of the weighted cost is improved through the use of iterative reweighted solution methods. Finally, the developed approach is summarized in a procedure.

\subsection{Linear Least-Squares for Local Rational Modeling Beyond the Nyquist Frequency}
\label{sec:SummedCost}
By appropriately weighting the non-linear cost function \eqref{eq:CostFunction}, a linear least-squares criterion is obtained, as shown in \lemmaRef{lemma:LinearizedSolution}.
\begin{lemma}
	\label{lemma:LinearizedSolution}
	By multiplying the residual $Y_l(k+r)-\widehat{Y}\left(k+r,\widehat{\Theta}(k)\right)$ in \eqref{eq:CostFunction} with
	\begin{equation}
		\label{eq:CommonDenominator}
		\begin{aligned}
			e(\freq_{k+r}) = 1+\sum_{s=1}^{R_e}e_s(k)r^s\equiv \prod_{f=0}^{\fac-1}d(\freq_{k+r+fM}),
		\end{aligned}
	\end{equation}
	resulting in the linear least-squares criterion
	\begin{equation}
		\label{eq:LinearCostFunction}
		\begin{aligned}
			J_{W}\left(\widetilde{\Theta}\left(k\right)\right) = \sum_{r=-\wsize}^{r=\wsize}\left\|
			Y_l(k+r)-\widetilde{\Theta}(k)K(k+r)
			\right\|_2^2.
		\end{aligned}
	\end{equation}
\end{lemma}
\begin{IEEEproof}
	Substituting $e(\freq_{k+r})$ in \eqref{eq:NonLinearAdditiveYlApprox} results in
	\begin{equation}
		\label{eq:NonLinearYlApprox}
		\resizebox{\linewidth}{!}{$ \displaystyle
			\displaystyle
			\begin{aligned}
				\widehat{Y}_l\!\left(\!k\!+\!r,\widetilde{\Theta}\!\left(k\right)\!\right) \!=& \frac{1}{\fac}\frac{\sum_{f=0}^{\fac-1}\!\left(\!\left(\widehat{G}\!\left(\freq_{k+fM}\!\right)\!+\!\sum_{s=1}^{R_g}g_s({k+fM})r^s\!\right)\!U(k+r+fM)\!\right)\!}{1+\sum_{s=1}^{R_e}e_s(k)r^s} \\
				&+ \frac{1}{\fac}\frac{\widehat{T}\left(\freq_k\right)+\sum_{s=1}^{R_t}t_s(k)r^s }{1+\sum_{s=1}^{R_e}e_s(k)r^s},
			\end{aligned}$}
	\end{equation}
	where 
	\begin{equation}
		\begin{aligned}
			R_g &=R_n+R_d(\fac-1),\\ R_t&=R_m+R_d(\fac-1),\\ R_e &= R_d\fac,
		\end{aligned}
	\end{equation}
	 result in the same model as in \eqref{eq:localPlant} and \eqref{eq:localTransient}. Furthermore, the numerator polynomials $g_s$ and $t_s$ are obtained by multiplying $n_s(\freq_{k+r+iM})$ and $m_s(\freq_{k+r+iM})$ from \eqref{eq:localPlant} and \eqref{eq:localTransient} with all $d(\freq_{k+r+fM}) \; \forall f\neq i$, i.e.,
	\begin{align}
		\label{eq:Gproduct} &\left(\widehat{G}(\freq_{k+iM})+\sum_{s=1}^{R_n}n_s({k+iM})r^s\right) \!\!\prod_{f\in\mathbb{Z}_{[0,\fac-1]\setminus i}}\!\!\!\!\!\!\! d(\freq_{k+r+fM})r^s \\
		\nonumber &\equiv \widehat{G}(\freq_{k+iM})+\sum_{s=1}^{R_g}g_s({k+iM})r^s, \\
		\label{eq:Tproduct}&\left(\widehat{T}(\freq_{k})+\sum_{s=1}^{R_m}m_s({k})r^s\right) \!\!\prod_{f\in\mathbb{Z}_{[0,\fac-1]\setminus i}}\!\!\!\!\!\!\! d(\freq_{k+r+fM})r^s \\
		\nonumber	&\equiv \widehat{T}(\freq_{k})+\sum_{s=1}^{R_t}t_s({k})r^s.
	\end{align}
	Then, by substituting \eqref{eq:NonLinearYlApprox} in the residual $Y_l(k+r)-\widehat{Y}\left(k+r,\widehat{\Theta}(k)\right)$ and multiplying with $e(\freq_{k+r})$ from \eqref{eq:CommonDenominator}, the linear least-squares criterion \eqref{eq:LinearCostFunction} is obtained.
\end{IEEEproof}
\begin{remark}
	LRM for single-rate LTI systems in the sense of \citet{McKelvey2012} is recovered by setting $\fac=1$.
\end{remark}
The linear least-squares criterion \eqref{eq:LinearCostFunction}, with output \eqref{eq:NonLinearYlApprox} and local models \eqref{eq:CommonDenominator}, \eqref{eq:Gproduct}, and \eqref{eq:Tproduct}, is formulated using the parameter row vector $\widetilde{\Theta}(k) \in\mathbb{C}^{1\times \left(R_g+1\right)\fac+R_t+1+R_e}$, i.e.,
\begin{equation}
	\label{eq:SummedParameters}
	\begin{aligned}
		\widetilde{\Theta}(k) = \begin{bmatrix}
			\theta_{\widehat{G}}\!\! & \theta_g & \widehat{T}(\freq_{k})\!\! & t_1({k}) & \!\!\!\cdots\!\!\! & t_{R_t}({k})\!\! &  \theta_e
		\end{bmatrix},
	\end{aligned}
\end{equation}
with $\theta_{\widehat{G}}$ from \eqref{eq:NonLinearParameters} and
	\begin{align}
		\theta_g &= \frac{1}{\fac}\begin{bmatrix}
			g_1(k) & g_1({k+M}) & \!\!\!\cdots\!\!\! & \cdots\!\!\! & g_{R_g}({k+(\fac-1)M})
		\end{bmatrix}, \nonumber\\
		\theta_e &= \hphantom{\frac{1}{\fac}}\begin{bmatrix}
			e_1(k) & e_2(k) &  \cdots & e_{R_e}(k)
		\end{bmatrix}.
	\end{align}
The input matrix $K(k+r)$ in \eqref{eq:LinearCostFunction} is given by
\begin{equation}
	\label{eq:Kvector}
	\begin{aligned}
		K(k+r) = \begin{bmatrix}
			K_1(r,R_g) \otimes \underline{U}(k+r) \\
			K_1(r,R_t) \\
			-K_2(r,R_d\fac) Y_l(k+r)
		\end{bmatrix},
	\end{aligned}
\end{equation}
with Kronecker product $\otimes$, $K_1(r,R) = \begin{bmatrix}	1 & r & \cdots & r^{R}\end{bmatrix}^\top$, $K_2(r,R) = \begin{bmatrix}	 r & \cdots & r^{R}\end{bmatrix}^\top$, and input vector
\begin{equation}
	\label{eq:liftedInput}
	\begin{aligned}
		\underline{U}(k+r) = \begin{bmatrix}
			U(k+r) \\
			U(k+r+M) \\
			\vdots \\
			U(k+r+(\fac-1)M)
		\end{bmatrix}.
	\end{aligned}
\end{equation}
%
The linear least-squares criterion \eqref{eq:LinearCostFunction} resolves the non-linear optimization challenges in \secRef{sec:NLMethod} with its closed-form minimizer, which is discussed next.
\subsection{Closed-Form Minimizer for Local Rational Modeling Beyond the Nyquist Frequency}
In this section, a closed-form minimizer to the linear least-squares criterion \eqref{eq:LinearCostFunction} is determined. The summation in \eqref{eq:LinearCostFunction} is removed by gathering the data in the window $r$ as
\begin{equation}
	\label{eq:GatheredLinearCostFunction}
	\begin{aligned}
		J_{W}\left(\widetilde{\Theta}\left(k\right)\right) = \left\|
		Y_{l,\wsize}-\widetilde{\Theta}(k) K_\wsize
		\right\|_2^2,
	\end{aligned}
\end{equation}
where $K_\wsize \in \mathbb{C}^{\left(R_g+1\right)\fac+R_t+1+R_e \times 2\wsize+1}$ and $Y_{l,\wsize} \in \mathbb{C}^{1\times 2\wsize+1}$  are constructed as
\begin{equation}
	\label{eq:gather}
	\begin{aligned}
		X_{\wsize}\!\!=\!\begin{bmatrix}
			X\left(k\!-\!\wsize\right) \!\!&\! X\left(k-\!\wsize\!+\!1\right) \!\!& \!\!\cdots\!\! &\!\! X\left(k\!+\!\wsize\right)
		\end{bmatrix}.
	\end{aligned}
\end{equation}
The minimizer to the cost function \eqref{eq:GatheredLinearCostFunction} leads to a least-squares closed-form solution for local rational modeling beyond the Nyquist frequency, that is
\begin{equation}
	\label{eq:linearSolution}
	\begin{aligned}
		\widetilde{\Theta}(k) = Y_{l,\wsize}K_\wsize^H\left(K_\wsize K_\wsize^H\right)^{-1}.
	\end{aligned}
\end{equation}
The fast-rate models at the frequency bands $k+fM \; \forall f\in\mathbb{Z}_{[0,\fac-1]}$ are jointly estimated by
\begin{equation}
	\label{eq:linearPlantSolution}
	\begin{aligned}
		\begin{bmatrix}
			\widehat{G}(\freq_{k}) \\ \widehat{G}(\freq_{k+M}) \\ \vdots \\ \widehat{G}(\freq_{k+(\fac-1)M})
		\end{bmatrix}^\top &= \fac \widetilde{\Theta}(k)\begin{bmatrix}
		I_\fac & 0_{\fac\times R_g\fac+R_t+1+R_e}
	\end{bmatrix}^\top\!\!, \\ 
&\forall k\in\mathbb{Z}_{[0,M-1]},
	\end{aligned}
\end{equation}
and similarly for the transient $\widehat{T}(\freq_{k})$.

Necessary conditions for the uniqueness of the closed-form solution \eqref{eq:linearSolution} to linear least-squares criterion $J_W$ in \eqref{eq:LinearCostFunction} are given by
	\begin{subequations}
	\label{eq:uniqueConditions}
	\begin{align}
		2\wsize+1&\geq \left(R_g+1\right)\fac+R_t+1+R_e, \label{eq:cond1} \\
		2\wsize+1&\leq M, \label{eq:cond2} \\
		\begin{split}
			\label{eq:cond3} 
			\big|U(k+r_1+i&M)-U(k+r_2+iM)\big|\neq 0, \\
			\forall r_1,r_2\in&\mathbb{Z}_{[-\wsize,\wsize]}, \;\forall i\in\mathbb{Z}_{[0,\fac-1]}.
		\end{split}
	\end{align}
\end{subequations}
%
In practice, it is observed that the conditions \eqref{eq:uniqueConditions} are also sufficient for the unique existence of the closed-form solution \eqref{eq:linearSolution}. The conditions \eqref{eq:uniqueConditions} are to be interpreted as design criteria on the input and local models as follows.
\begin{enumerate}
	\item For each frequency bin $k$, the amount of estimated parameters $\left(R_g+1\right)\fac+R_t+1+R_e$ in $\widetilde{\Theta}$ should be less than the number of data points $2\wsize+1$, leading to \eqref{eq:cond1}. This intuitively explains how the local models with $\left(R_g+1\right)\fac+R_t+1+R_e$ parameters $\widetilde{\Theta}$ in \eqref{eq:linearSolution} enable disentangling \fac aliased contributions through the use of $2\wsize+1$ outputs $Y_l(k+r)$.	
	\item The window size $2\wsize+1$ should not exceed the amount of data points $M$ in $Y_{l,\wsize}$, leading to \eqref{eq:cond2}.
	\item $K_{\wsize}$ is full (row) rank if all inputs in the local and aliased windows $\underline{U}(k+r)$ are sufficiently 'rough', which is formalized for a single local window in \citet{Schoukens2009}. For \eqref{eq:linearSolution}, this implies that the spectral difference $\left|U(k+r_1+iM)-U(k+r_2+iM)\right|$ in \eqref{eq:cond3} should not become zero, which is fulfilled by for example random-phase multisines.
\end{enumerate}

\begin{remark}
	To prevent the Vandermonde structure in the matrix $K_\wsize$ from becoming ill-conditioned, $R_g$, $R_t$, and $R_e$ should not be chosen excessively high. Alternatively, the numerical conditioning can be improved according to \citet{Voorhoeve2018}.
\end{remark}
\begin{remark}
	Identification of FRFs beyond the Nyquist frequency using slow-rate outputs with LPM \citep{VanHaren2023a} is a special case of the developed framework by setting $R_e=0$ and $R_g=R_t$.
\end{remark}
\begin{remark}
	The windows for the left and right frequency borders are $k+r\in \mathbb{Z}_{[0,2\wsize]} \; \forall k \leq \wsize$ and $k+r\in\mathbb{Z}_{[M-2\wsize,M]} \; \forall k>M-\wsize$, similarly to \citet[Section~7.2.2.6]{Pintelon2012}.
\end{remark}

Additionally, the closed-form solution enables approximating the variance, which is presented in \lemmaRef{lemma:VarianceLinearLRM}.
\begin{lemma}
	\label{lemma:VarianceLinearLRM}
	Let the system and transient be modeled using \eqref{eq:localPlant} and \eqref{eq:localTransient}. Then, the estimated variance of the FRF $\widehat{G}(\freq_k)$, determined with \eqref{eq:linearPlantSolution}, is given by
	\begin{equation}
		\label{eq:Variance}
		\begin{aligned}
			\mathrm{var} \left(\widehat{G}\left(\freq_k\right)\right) \approx \fac^2 \overline{S^H S}\widehat{C}_v(k) \; \forall k \in\mathbb{Z}_{[0,M-1]},
		\end{aligned}
	\end{equation}
	that is an estimate of the true variance of the identified FRF
	\begin{equation}
		\begin{aligned}
			\mathrm{var} \left(\widehat{G}(\freq_k)\right) = \fac^2 \overline{\mathbb{E}\left(S^H S\right)}{C}_v(k)+\fac^2 O_{\mathrm{int,H}}\left(\frac{n_w^0}{M}\right),
		\end{aligned}
	\end{equation}
	with variance of the noise $C_v$ and its estimate $\widehat{C}_v$, noise interpolation error $O_{\mathrm{int,H}}$ \citep{Pintelon2012}, and
	\begin{equation}
		\begin{aligned}
			S=K_\wsize^H \left(K_\wsize K_\wsize^H\right)^{-1} \begin{bmatrix}1 & 0_{1\times\fac(R_g+1)+R_t+R_e} \end{bmatrix},
		\end{aligned}
	\end{equation}
and similarly for the FRF at the frequency bands $k+fM \; \forall f\in\mathbb{Z}_{[1,\fac-1]}$.
\end{lemma}
\begin{IEEEproof}
	The results of \citet[Appendix~7.E]{Pintelon2012} and \citet{VanHaren2023a} apply respectively for $\fac=1$, and $R_e=0,\;R_g=R_t$. Furthermore, for arbitrary $\fac,$ $R_e,$ $R_g$ and $R_t$, the system output for window \wsize{} for the local models in \eqref{eq:localPlant} and \eqref{eq:localTransient} is given by
	\begin{equation}
		\label{eq:trueSummedOutput}
		\begin{aligned}
			Y_{l,\wsize} = \widetilde{\Theta}_0(k)K_\wsize+V_{l,\wsize},
		\end{aligned}
	\end{equation}
	where $\widetilde{\Theta}_0(k)$ has the same structure as $\widetilde{\Theta}(k)$ in \eqref{eq:SummedParameters}, but containing the true parameters of the system. Rewriting \eqref{eq:trueSummedOutput}, multiplying by $S$, and combining with \eqref{eq:linearSolution} results in
	\begin{equation}
		\begin{aligned}
			\left(Y_{l,\wsize}-\widetilde{\Theta}_0(k)K_\wsize\right)S = V_\wsize S&, \\
			\widetilde{\Theta}(k)\begin{bmatrix} 1 & 0 	\end{bmatrix}^\top  - \widetilde{\Theta}_{0}(k) \begin{bmatrix} 1 & 0 	\end{bmatrix}^\top= V_\wsize S&, \\
			\frac{1}{\fac}\widehat{G}(\freq_k)-\frac{1}{\fac}G(\freq_k) = V_\wsize S&, \\
			\widehat{G}(\freq_k) = G(\freq_k)+\fac V_\wsize S&,
		\end{aligned}
	\end{equation}
	where the variance is calculated as $\textrm{var}\left(\widehat{G}(\freq_k)\right) = \mathbb{E}\left(\widehat{G}(\freq_k)\widehat{G}^H(\freq_k)\right)$.
\end{IEEEproof}

The variance of the noise ${C}_v(k)=\textrm{var}\left(V_l(k)\right)=\mathbb{E}\left(V_l(k)V_l^H(k)\right)$ is estimated using the residual of the least-squares fit \citep[Appendix~7.B]{Pintelon2012}, i.e.,
\begin{equation}
	\begin{aligned}
		\widehat{C}_v(k) = \frac{1}{q}\left(Y_{l,\wsize}\!\!-\widetilde{\Theta}(k)K_\wsize \right)\!\!\left(Y_{l,\wsize}\!\!-\widetilde{\Theta}(k)K_\wsize\right)^H\!,
	\end{aligned}
\end{equation}
with degrees of freedom $q=2\wsize+1-\left(\left(R_g+1\right)\fac+R_t+1+R_e\right)$.


The linear least-squares criterion \eqref{eq:LinearCostFunction} is obtained by appropriately weighting the non-linear cost function in \eqref{eq:CostFunction}, which has closed-form solution \eqref{eq:linearSolution}. Generally, this method is effective, especially for practical applications \citep{Voorhoeve2018,Verbeke2020}. Additionally, iterative reweighted solutions further enhance the accuracy of weighted linear least-squares.
\subsection{Iterative Reweighted Solutions}
\label{sec:IterativeRefinement}
The accuracy of the weighted least-squares criterion \eqref{eq:LinearCostFunction} is improved through iterative reweighted solutions. Both the Sanathanan-Koerner (SK) and Levenberg-Marquardt (LM) algorithms are employed for this purpose.
\paragraph{Sanathanan-Koerner Algorithm}
The SK algorithm \citep{Sanathanan1963} iteratively counteracts the weighting from \eqref{eq:LinearCostFunction} by reweighting with its inverse determined in the previous SK iteration. Hence, the iteratively minimized cost function is
\begin{equation}
	\label{eq:LinearizedCostFunctionSK}
	\begin{aligned}
		&J_{SK}\left(\widetilde{\Theta}^{<j>}\left(k\right)\right)= \!\!\sum_{r=-\wsize}^{\wsize}\Bigg\|
		\Bigg(1+\sum_{s=1}^{R_e}e_s^{<j-1>}(k)r^s\Bigg)^{-1} \\
		&\Bigg(1+\sum_{s=1}^{R_e}e_s^{<j>}(k)r^s\Bigg)\!\!\Bigg(Y_l(k+r) - \widehat{Y}_l(k+r,\widetilde{\Theta}^{<j>}(k))\Bigg)
		\Bigg\|_2^2\!\!,
	\end{aligned}
\end{equation}
which is minimized until convergence or a stopping criterion is met. As initial guess, the closed-form solution from \eqref{eq:linearSolution} can be used, i.e.,
\begin{equation}
	\widetilde{\Theta}^{<0>}(k) = \widetilde{\Theta}(k) \;\;\forall k.
\end{equation}
Conveniently, the iterative solution is determined similar to the closed-form solution in \eqref{eq:linearSolution}, that is,
\begin{equation}
	\begin{aligned}
		\widetilde{\Theta}^{<j>}(k) = Z_{l,\wsize}^{<j-1>}\left(\vphantom{\left(L_\wsize^{<j-1>}\right)^H}L_\wsize^{<j-1>}\right)^H\!\left(L_\wsize^{<j-1>} \left(L_\wsize^{<j-1>}\right)^H\right)^{-1}\!\!\!\!,
	\end{aligned}
\end{equation}
where $Z_{l,\wsize}$ and $L_\wsize$ are constructed according to \eqref{eq:gather}, with their components 
\begin{equation}
	\begin{aligned}
		Z_{l}^{<j-1>}(k+r) &= 	\left(1+\sum_{s=1}^{R_e}e_s^{<j-1>}(k)r^s\right)^{-1} Y_l(k+r), \\
		L^{<j-1>}(k+r) & = \left(1+\sum_{s=1}^{R_e}e_s^{<j-1>}(k)r^s\right)^{-1} K(k+r).
	\end{aligned}
\end{equation}
The SK algorithm has been successfully applied in the system identification literature with attractive convergence properties, specifically in practical situations \citep{Friedman1995}. The accuracy of the solution with respect to the original cost function is further increased via the LM algorithm.
\paragraph{Levenberg-Marquardt Algorithm}
Second, a gradient-based non-linear optimizer can be used to optimize the original cost function \eqref{eq:CostFunction} with parameters $\widetilde{\Theta}$, i.e.,
\begin{equation}
	\begin{aligned}
		J_{LM}\left(\widetilde{\Theta}(k)\right) &= 	J_{LS}\left(\widehat{\Theta}(k)\right) \\
		&= \sum_{r=-\wsize}^{\wsize}\left\|
		Y_l(k+r) - \widehat{Y}_l\left(k+r,\widetilde{\Theta}\left(k\right)\right)
		\right\|_2^2,
	\end{aligned}
\end{equation}
where $\widehat{Y}_l\left(k+r,\widetilde{\Theta}\left(k\right)\right)$ is calculated using \eqref{eq:NonLinearYlApprox}. The LM algorithm, which is a damped Newton-Gauss algorithm, has been successfully applied for local modeling in \citet{Voorhoeve2018}. Until convergence or a stopping criterion is met, the parameters are updated based on the gradient of the cost function with respect to the parameters. As initial guess, the closed-form solution from \eqref{eq:linearSolution} can be used. Alternatively, the LM algorithm can be used complementary to the SK algorithm by using the result from the SK algorithm as initial guess. The LM algorithm optimizes the original cost function, and hence, a local minimum of the original cost function is found. However, since the cost function is non-linear, the LM algorithm does not guarantee convergence to the global minimum, and its result is strongly dependent on the initial guess.

With this in mind, it is recommended to start with solution \eqref{eq:linearSolution} of the linear least-squares criterion \eqref{eq:LinearCostFunction} and perform SK iterations. Subsequently, the result of the SK algorithm serves as a good initial guess for the LM algorithm.
\subsection{Procedure for Local Rational Modeling Beyond The Nyquist Frequency}
The main results in Sections~\ref{sec:NLMethod} and \ref{sec:LinearMethod} are summarized in Procedure~\ref{proc:1}. 
\begin{figure}[H]
	\vspace{6pt}\hrule \vspace{1mm}\begin{proced}[Identify fast-rate FRF using slow-rate outputs and fast-rate broadband inputs with local rational model] \hfill \vspace{0.5mm} \hrule \vspace{1mm}
		\label{proc:1}
		\begin{enumerate}
			\item Construct $u_h$, such that it satisfies the requirements of $\underline{U}$ in \eqref{eq:cond3}, e.g., random-phase multisines.
			\item Apply input $u_h$ to system and record the output $y_l$.
			\item Take the DFT of input $u_h$ and output $y_l$ using \eqref{eq:DFT3}.
			\item For frequency bins $k\in\mathbb{Z}_{[0,M-1]}$, determine $\widehat{G}(\freq_{k+fM}) \;\forall f\in\mathbb{Z}_{[0,\fac-1]}$, which can be done in the following two ways.
			\begin{enumerate}
				\item Directly optimize the non-linear cost in \eqref{eq:CostFunction}, using a non-linear optimizer.
				\item Or, optimize the linear least-squares criterion \eqref{eq:LinearCostFunction}.
				\begin{enumerate}
					\item Use \eqref{eq:gather} to construct matrices $K_{\wsize}$ and $Y_{l,\wsize}$, with measured outputs $Y_l(k+r)$ and input vectors $K(k+r)$ from \eqref{eq:Kvector}, using $\underline{U}(k+r)$.
					\item Compute parameter vector $\widetilde{\Theta}(k)$ from \eqref{eq:linearSolution}, and consequently the FRF $\widehat{G}\left(\freq_{k+fM}\right)\;\forall f\in\mathbb{Z}_{[0,\fac-1]}$ using \eqref{eq:linearPlantSolution}.
					\item Calculate the estimated variance of the FRF with \lemmaRef{lemma:VarianceLinearLRM}.
					\item Optional: Refine optimizer using the SK or LM algorithms, as described in \secRef{sec:SummedCost}.
				\end{enumerate}
			\end{enumerate}
		\end{enumerate}
		\vspace{0pt} 	\hrule \vspace{-9pt}
	\end{proced}
\end{figure}
\section{Experimental Validation}
\label{sec:example}
In this section, the developed framework is validated on a prototype wafer stage used for semiconductor manufacturing, leading to contribution C3. The experimental setup is introduced, followed by the results. Finally, the FRF is refined using the iterative procedures from \secRef{sec:IterativeRefinement}.
\subsection{Experimental Setup}
\begin{figure}[tb]
	\centering
	\includegraphics{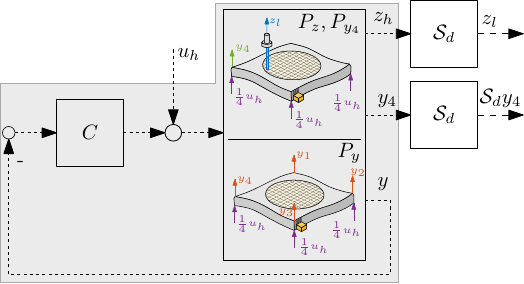}
	\caption{Experimental feedback scheme used, where the equivalent systems in \eqref{eq:EquivalentSystems} are to be identified.}
	\label{fig:ExperimentalScheme}
\end{figure}

The experimental setup is the OAT shown in \figRef{fig:ExperimentalSetup}. The chuck is levitated and actuated by four Lorentz type actuators on the corners of the chuck. Additionally, the vertical displacement is measured at the corners of the chuck by means of 4 linear encoder heads and scales, and are used as inputs to an internal feedback controller as $y=\frac{1}{4}\left(y_1+y_2+y_3+y_4\right)$. For the case study, the fast-rate excitation $u_h$ is equally distributed over the four actuators, and is considered a disturbance to the plant. The scanning sensor is suspended above the wafer, and can be moved in the horizontal plane. The scanning sensor is positioned in the bottom left corner of the wafer, 110 mm in both directions from the center. A schematic overview of the control scheme is seen in \figRef{fig:ExperimentalScheme}. 

The goal is to identify fast-rate (equivalent) models using slow-rate outputs. Specifically, FRFs are identified for the displacement of the point-of-interest $z_h$, in addition to one of the corners of the chuck $y_4$, i.e.,  
\begin{equation}
	\label{eq:EquivalentSystems}
	\begin{aligned}
		G_{y_4}\left(\freq_k\right) &= P_{y_4}\left(\freq_k\right)\left(I+C\left(\freq_k\right)P_y\left(\freq_k\right)\right)^{-1}: u_h\mapsto y_4, \\
		G_z\left(\freq_k\right) &= P_z\left(\freq_k\right)\left(I+C\left(\freq_k\right)P_y\left(\freq_k\right)\right)^{-1}: u_h\mapsto z_h,
	\end{aligned}
\end{equation}
are identified using the fast-rate input $u_h$ and slow-rate outputs $\mathcal{S}_dy_4$ and $z_l$.
The excitation signal is a single period of a random-phase multisine, exciting all frequencies with a flat amplitude spectrum, and having a root mean square value of 1.44 N. The signal-to-noise ratio, which is the ratio of the variance of the output $z_l$ to the noise $v_l$, is estimated to be around 45 dB. Further experimental settings are shown in \tabRef{tab:expValues}.
\begin{table}[tb]
	\centering
	\caption{Experimental settings.}
	\label{tab:expValues}
	\begin{tabular}{lll}
		\toprule
		\textbf{Variable}    & \textbf{Abbrevation} & \textbf{Value}  \\
		\midrule
		Fast sampling time   & \tsh            & 0.5 ms       \\
		Slow sampling time & \tsl 				& 1.5 ms	 \\
		Downsampling factor  & \fac            & 3           \\
		Number of input samples    & N               & 1200      \\
		Number of output samples    & M              & 400      \\
		Measurement time & $T_m$ & 0.6 s \\
		System numerator degree & $R_g$ & 4  \\
		Transient numerator degree & $R_t$ & 4 \\
		Denominator degree & $R_e$ & 7  \\
		Window size & \wsize & 18  \\
		\bottomrule
	\end{tabular}
\end{table}
The following methods are compared.
\begin{enumerate}[align=left, labelsep=0.4cm, itemindent = 0.2cm]
	\item[$\widehat{G}_{LRM}$] The developed approach using the closed-form solution \eqref{eq:linearSolution}, with settings as shown in \tabRef{tab:expValues}. 
	\item[$\widehat{G}_{LPM}$] The approach from \citet{VanHaren2023a}, i.e., the developed approach with $R_d=R_e=0$, $R_g\equiv R_t = 2$, and $\wsize=18$.
	\item[$\widehat{G}_{SA}$] A traditional approach using Spectral Analysis (SA) with a Hanning window.
\end{enumerate}
The SA assumes that the DFT is periodic in the slow sampling frequency, i.e., $\widehat{Y}_h(\freq_{k+iM}) = \fac Y_l(\freq_{k}) \:\forall i\in\mathbb{Z}$, which in time-domain is equivalent to interpolating $z_l$ with zeros as $\widehat{z}_h = \begin{bmatrix}	z_l(0) & 0  & 0 & z_l(1) & \cdots \end{bmatrix}^\top$, and similarly for $y_4$. SA averages the FRF over 11 segments of the data as
\begin{equation}
	\label{eq:hann}
	\begin{aligned}
		\widehat{G}_{SA}(\Omega_p) =\frac{\sum_{i=1}^{11}\widehat{Y}_{h,i}(\Omega_p)\overline{U}_{h,i}(\Omega_p)}{\sum_{i=1}^{11}{U}_{h,i}(\Omega_p)\overline{U}_{h,i} (\Omega_p)} \:\: \forall p \in \{0,6,12,\ldots\},
	\end{aligned}
\end{equation}
where ${X}_i$ is the DFT of the $i^{\textrm{th}}$ segment of $X$, multiplied with a Hanning window, where each segment contains 200 samples with 100 samples overlap. For validation purposes, the outputs $z_l$ and $\mathcal{S}_d y_4$ are recorded at the fast sampling rate as well, i.e., $z_h$ and $y_4$ are available, and an FRF is constructed using $N=50000$ samples with the local rational modeling method from \citet{McKelvey2012}, with rational degrees $R_n=R_d=R_m=4$ and a window size of $\wsize=150$. The cumulative FRF error for the first $n$ frequencies is defined as
\begin{equation}
	\label{eq:CPSD}
	\begin{aligned}
		\frac{1}{N} \sum_{k=1}^{n} \left|
		G(\freq_k)-\widehat{G}(\freq_k)
		\right|.
	\end{aligned}
\end{equation}
\subsection{Experimental Results}
\label{sec:ExpResults}
\begin{figure*}[tb]
	\centering
	\includegraphics{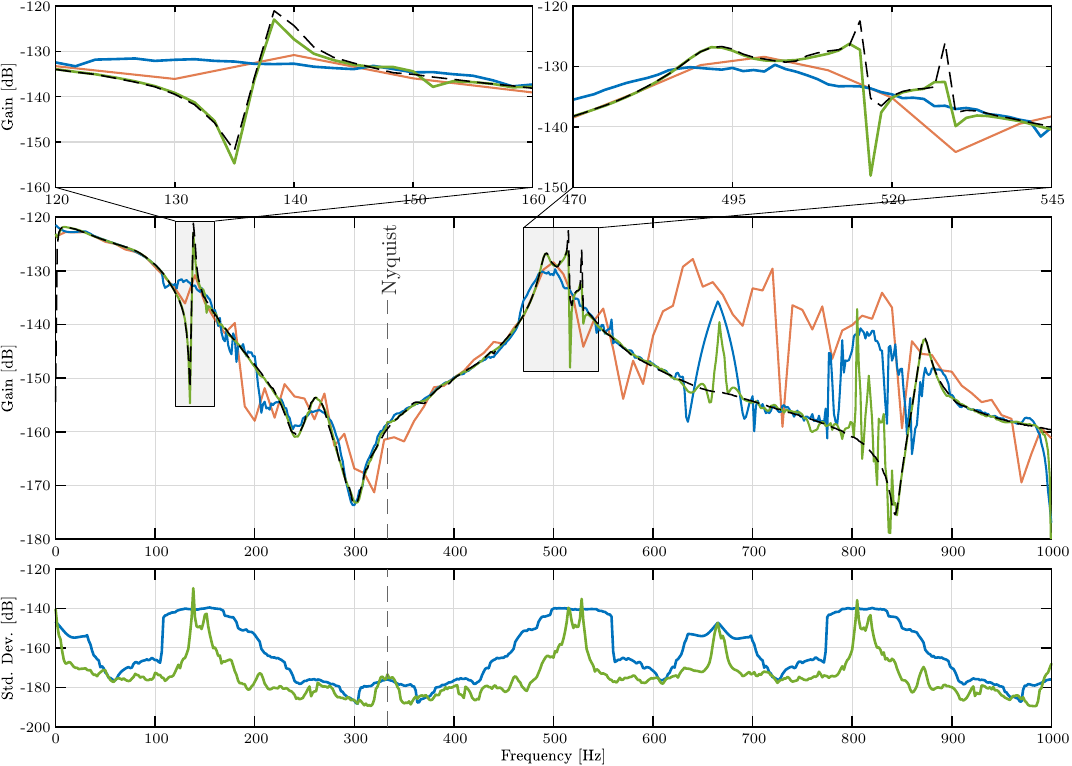}
	\caption{(Middle:) The developed local rational modeling method $\widehat{G}_{LRM}(\Omega_k)$ \markerline{mgreen} identifies the true system $G_{y_4}(\Omega_k)$ \markerline{black}[densely dashed][*][0][0.8] accurately, even beyond the Nyquist frequency \markerline{gray}[densely dashed][^][0][0.5] and the lightly-damped resonant dynamics (enlarged in top). Both $\widehat{G}_{LPM}(\Omega_k)$ \markerline{mblue} and $\widehat{G}_{SA}(\Omega_k)$ \markerline{mred} identify the true system $G(\Omega_k)$ \markerline{black}[densely dashed][*][0][0.8] significantly less accurate. (Bottom:) The associated standard deviation for $\widehat{G}_{LPM}(\Omega_k)$ \markerline{mblue} and $\widehat{G}_{LRM}(\Omega_k)$ \markerline{mgreen}, calculated using the square root of \eqref{eq:Variance}.}
	\label{fig:OATExpNodal}
\end{figure*}

\begin{figure*}[tb]
	\centering
	\includegraphics{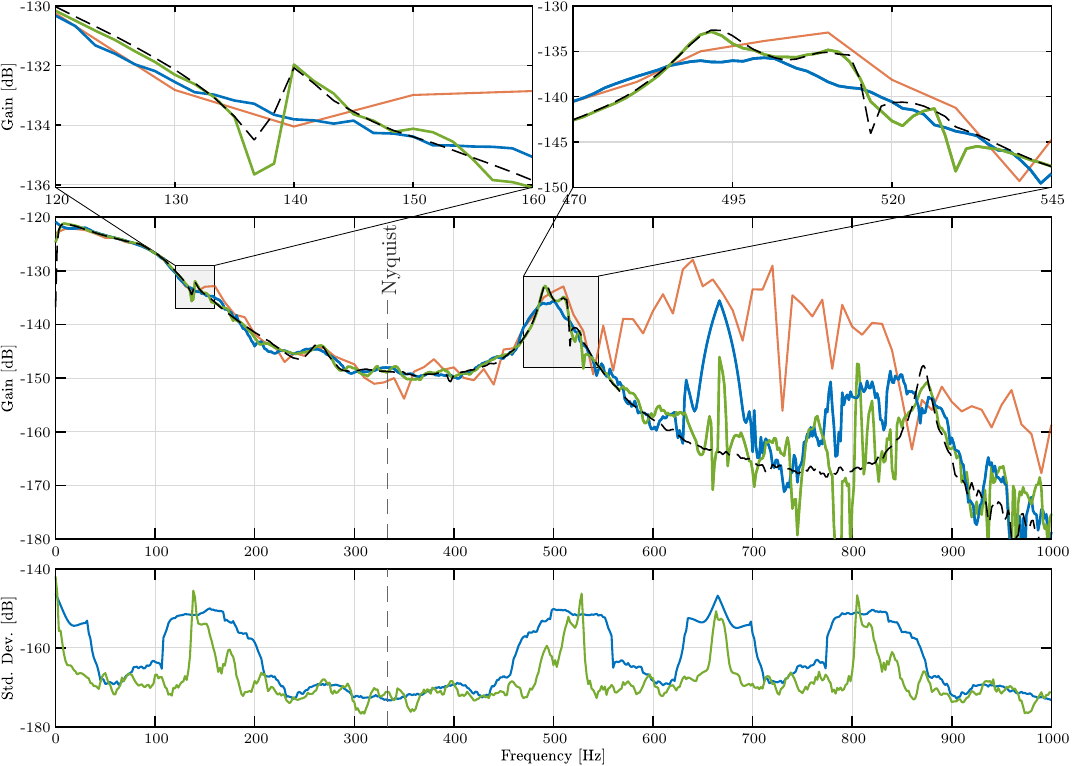}
	\caption{(Middle:) The developed local rational modeling method $\widehat{G}_{LRM}(\Omega_k)$ \markerline{mgreen} identifies the true system $G_z(\Omega_k)$ \markerline{black}[densely dashed][^][0][0.8] accurately, even beyond the Nyquist frequency \markerline{gray}[densely dashed][^][0][0.5] and the lightly-damped resonant dynamics (enlarged in top). Both $\widehat{G}_{LPM}(\Omega_k)$ \markerline{mblue} and $\widehat{G}_{SA}(\Omega_k)$ \markerline{mred} identify the true system $G(\Omega_k)$ \markerline{black}[densely dashed][^][0][0.8] significantly less accurate. (Bottom:) The associated standard deviation for $\widehat{G}_{LPM}(\Omega_k)$ \markerline{mblue} and $\widehat{G}_{LRM}(\Omega_k)$ \markerline{mgreen}, calculated using the square root of \eqref{eq:Variance}.} 
	\label{fig:OATExpsss}
\end{figure*}

\begin{figure}[tb]
	\centering
	\includegraphics{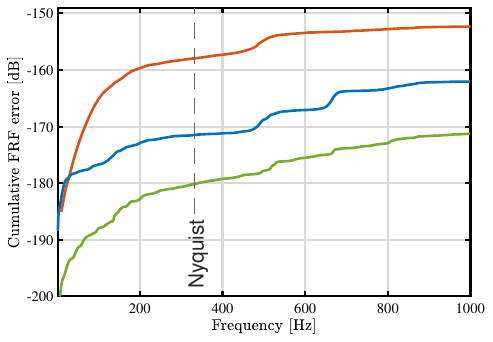}
	\caption{The developed LRM method \markerline{mgreen} achieves the lowest cumulative FRF error from \eqref{eq:CPSD} for $G_z\left(\freq_k\right)$ compared to the LPM method from \citet{VanHaren2023a} \markerline{mblue} and the SA method \eqref{eq:hann} \markerline{mred}.}
	\label{fig:CPSDFRFError}
\end{figure}

The true and estimated FRFs of $G_{z_4}$ and $G_y$ are seen in \figRef{fig:OATExpNodal} and \figRef{fig:OATExpsss}, with the cumulative error \eqref{eq:CPSD} of $G_y$ in \figRef{fig:CPSDFRFError}.
The following observations are made.
\begin{itemize}
	\item From the FRFs of $G_{y_4}$ and $G_z$ in respectively \figRef{fig:OATExpNodal} and \figRef{fig:OATExpsss}, the following is observed.
	\begin{itemize}
		\item The developed method $\widehat{G}_{LRM}$ estimates FRFs $G_{y_4}$ and $G_z$ accurately, even beyond the Nyquist frequency and including resonances and anti-resonances.
		\item The LPM method $\widehat{G}_{LPM}$ estimates the FRFs of $G_{y_4}$ and $G_z$ adequately. However, the lightly-damped resonant dynamics are not captured accurately due to the polynomial model structure.
		\item The SA method $\widehat{G}_{SA}$ estimates the FRFs $G_{y_4}$ and $G_z$ poorly, resonances and anti-resonances are modeled incorrectly, specifically above the Nyquist frequency. Additionally, the frequency resolution is a factor 6 lower compared to the LPM and LRM method.
		\item The developed LRM method results in a significantly lower standard deviation compared to the LPM approach, particularly around resonances and their aliased frequencies. This is expected, as rational models are more suitable for lightly-damped resonant dynamics.
	\end{itemize}
	\item \figRef{fig:CPSDFRFError} clearly shows that the LRM method achieves the lowest cumulative error of $G_z$.
\end{itemize}
The observations show that the LRM method is most suitable for identifying fast-rate FRFs of (lightly-damped) systems using slow-rate outputs. The LRM method is effective because it disentangles the aliased components through local models, and the rational model structure is capable of estimating the (lightly-damped) resonant dynamics accurately.

\subsection{Iterative Analysis}
\begin{figure}[tb]
	\centering
	\includegraphics{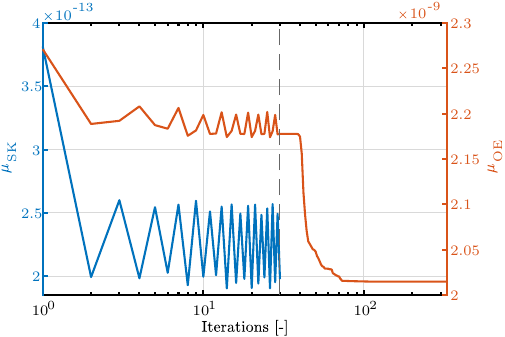}
	\caption{Mean SK \markerline{mblue}[solid] (left axis) and  mean least-squares cost \markerline{mred}[solid] from \eqref{eq:meanCosts} when estimating $G_{y_4}$ (right axis). After 30 SK iterations \markerline{gray}[densely dashed][o][0][0.5], an additional 300 iterations are done based on the LM algorithm per frequency.}
	\label{fig:SKCost}
\end{figure}

\begin{figure*}[t]
	\centering
	\includegraphics{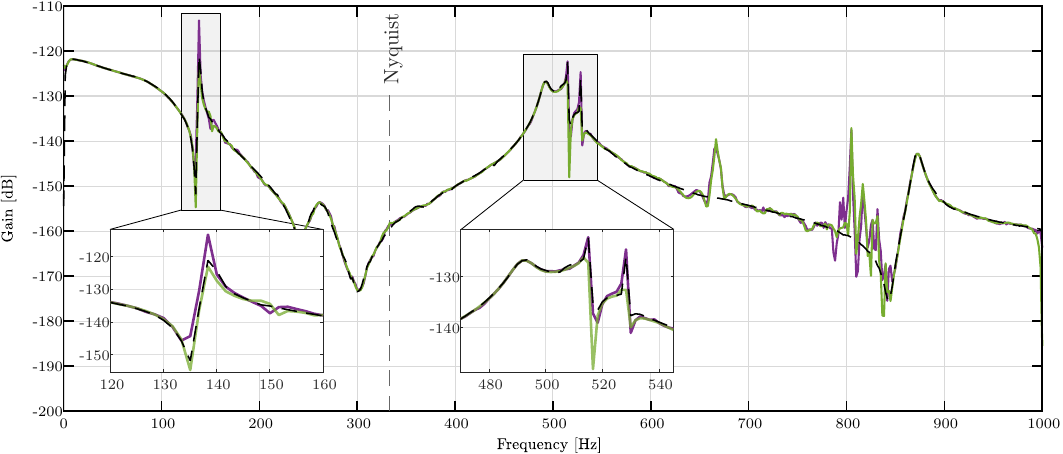}
	\caption{The FRF determined with closed-form solution \eqref{eq:linearSolution} \markerline{mgreen} and the FRF after 30 SK and 300 LM iterations \markerline{mpurple} both identify the true FRF $G_{y_4}\left(\freq_{k}\right)$ \markerline{black}[densely dashed][*][0][0.8] accurately.}
	\label{fig:OAT_Exp_SSS_Refinement}
\end{figure*}

The iterative reweighted solutions described in \secRef{sec:IterativeRefinement} are analyzed for the experimental validation. The mean SK cost and mean least-squares cost for all frequencies are defined as
\begin{equation}
	\label{eq:meanCosts}
	\begin{aligned}
		&\mu_{SK} = \frac{1}{N}\sum_{k=0}^{N-1} J_{SK}\left(\widetilde{\Theta}(k)\right), \\
		&\mu_{OE} = \frac{1}{N}\sum_{k=0}^{N-1}J_{LS}\left(\widetilde{\Theta}\left(k\right)\right),
	\end{aligned}
\end{equation}
where $J_{SK}$ and $J_{LS}$ are calculated with \eqref{eq:LinearizedCostFunctionSK} and \eqref{eq:CostFunction}, with estimated output $\widehat{Y}_l$ from \eqref{eq:NonLinearAdditiveYlApprox}. The mean SK and least-squares cost and the FRF after 30 SK and 300 LM iterations are seen in \figRef{fig:SKCost} and \figRef{fig:OAT_Exp_SSS_Refinement}.
The following observations are made with respect to iterative reweighting the experimental FRF.
\begin{itemize}
	\item Both the mean SK and mean least-squares cost in \figRef{fig:SKCost} are decreasing, as expected.
	\item After iterative reweighting, the resonant behavior around 500 Hz is estimated slightly more accurate compared to the closed-form solution, as observed from \figRef{fig:OAT_Exp_SSS_Refinement}.
	\item The mean least-squares cost in \figRef{fig:SKCost} decreases by almost 12\% after 30 SK and 300 LM iterations. On the other hand, the FRF after 30 SK and 300 LM iterations in \figRef{fig:OAT_Exp_SSS_Refinement} shows no significant difference compared to the FRF of the closed-form solution, indicating that iterative reweighting is not strictly necessary.
\end{itemize}
It is concluded that the weighted linear least-squares \eqref{eq:LinearCostFunction} is suitable for identifying fast-rate FRFs beyond the Nyquist frequency of a slow-rate output of the OAT. This can be explained because weighting the cost function only has a minor effect in the local windows, which was also observed in \citet{Voorhoeve2018}.

\section{Conclusions}
The results in this \manuscript enable identifying fast-rate FRFs where aliasing occurs due to slow-rate outputs. The key idea is to parameterize the system and transient FRFs through multiple local rational models, which allow to appropriately disentangle aliased contributions when exciting the full frequency spectrum. The local rational models are effective in modeling rational systems, especially with lightly-damped resonant dynamics. A linear least-squares criterion with a unique closed-form solution is determined by appropriately weighting the associated non-linear cost function. The closed-form solution does not suffer from local minima, in contrast to non-linear optimization, and additionally enables estimating the variance of the identified FRF. Furthermore, the estimation accuracy of the weighted linear least-squares is improved by means of iterative reweighted solutions, including the Sanathanan-Koerner algorithm. Finally, the framework is validated through experiments on a prototype wafer stage, demonstrating accurate identification of lightly-damped resonant dynamics beyond the Nyquist frequency. The developed approach plays a crucial role in identification and control design of closed-loop, multivariable, and parametric systems, especially for systems with slow-rate outputs, such as vision-in-the-loop systems.
\section*{Acknowledgement}
The authors would like to thank Leonid Mirkin for his help and fruitful discussions that led to the results in this paper.
\printbibliography
\end{document}